\documentclass[mathleft
]{an}

\usepackage{graphicx}
\usepackage{times}

\def\note #1]{{\bf #1]}}
\def\du{{\bf d}}
\def\ru{\mbox{\boldmath$r$}}
\def\xu{{\bf x}}
\def\Ku{{\bf K}}
\def\dqu{{\bf\delta q}}
\def\CKu{\mbox{\boldmath${\cal K}$}}

\def\id{{\rm d}}

\def\br{\mbox{\boldmath$r$}}
\def\bnabla{\mbox{\boldmath$\nabla$}}

\def\bS{\mbox{\boldmath$S$}}
\def\bxi{\mbox{\boldmath$\xi$}}
\def\los{\mbox{\boldmath$\ell$}}

\def\bdelta{\mbox{\boldmath$\Delta$}}

\def\bB{\mbox{\boldmath$B$}}
\def\bJ{\mbox{\boldmath$J$}}
\def\bU{\mbox{\boldmath$U$}}

\def\br{{\mbox{\boldmath$r$}}}
\def\bk{\mbox{\boldmath$k$}}

\def\id{{\rm d}}

\def\1{\mbox{\boldmath$x$}_1}
\def\2{\mbox{\boldmath$x$}_2}
\newcommand{\bx}{{\mbox{\boldmath$x$}}}

\begin{document}

\Pagespan{204}{}
\Yearpublication{2007}%
\Month{3}%
\Volume{328}%
\Issue{3-4}%
\DOI{10.1002/asna.200610720}%

\title{Outstanding problems in local helioseismology}
\author{L. Gizon\inst{1}\fnmsep\thanks{Corresponding author:
  \email{gizon@mps.mpg.de}\newline}
\and M.~J. Thompson\inst{2}  
}
\titlerunning{Outstanding Problems in Local Helioseismology}
\authorrunning{Gizon \& Thompson}
\institute{{Max-Planck-Institut f\"ur Sonnensystemforschung, 37191 Katlenburg-Lindau, Germany}
\and
{Department of Applied Mathematics, University of Sheffield, Sheffield S32~2HE, U.K.}
}

\received{24 Dec 2006}
\accepted{1 Jan 2007}
\publonline{later}


\abstract{
Time-distance helioseismology and related techniques show great promise for probing the structure and dynamics of the subphotospheric layers of the Sun. Indeed time-distance helioseismology has already been applied to make inferences about structures and flows under sunspots and active regions, to map long-lived convective flow patterns, and so on. Yet certainly there are still many inadequacies in the current approaches and, as the data get better and the questions we seek to address get more subtle, methods that were previously regarded as adequate are no longer acceptable. Here we give a short and partial description of outstanding problems in local helioseismology, using time-distance helioseismology as a guiding example.}

\maketitle

\sloppy

\section{Introduction}

Local helioseismology has grown spectacularly in the past decade. Global
helioseismology, using principally the frequencies of global modes, has
enabled us to image the radial variation of the hydrostatic structure
through most of the solar interior, and similarly the radial and latitudinal 
variation of the rotation rate. But some interesting features 
of the solar activity cycle -- sunspots, active regions --  are 
by definition horizontally localized and influence global-mode
frequencies only very subtly. Global modes are not the best instruments
to probe these features. Local helioseismology is much better suited to 
this task. A number of local helioseismological techniques and approaches
have been developed. These include time-distance helioseismology
(Duvall et al. 1993), ring analysis (Hill 1988; Antia \& Basu 2007), acoustic holography
(Lindsey \& Braun 2000) and statistical waveform analysis (Woodard 2002). In this
paper we are concerned mostly with time-distance helioseismology: for 
a recent review, see Gizon \& Birch (2005).
 
Time-distance helioseismology proceeds by cross-correlating 
measurements (typically Doppler velocities) of the oscillations at
different locations on the solar surface, fitting the cross-correlation
to determine a `travel time' for waves propagating between those different
locations, and then interpreting those travel times via some inversion
procedure to infer something about the subsurface conditions affecting
the wave propagation. What issues might be addressed using helioseismology
regarding, for example, sunspots? The list includes the structure of a 
sunspot both horizontally and with depth, the flows around the 
sunspot and how these vary with depth and distance from the sunspot. 
How can we decide what is the best
technique to employ to exploit the data at hand?
Time-distance helioseismology has proved attractive for a number of
reasons. It is quite intuitive. It is fairly robust, in that the
main measurement is a fitted travel time rather than more detailed
aspects of the cross-correlation function or wave field which could
be more sensitive to the manner of mode excitation. It achieves a 
much better resolution than has hitherto been achieved by ring analysis,
a technique which requires observations taken over a relatively
large patch of Sun in order to get adequate wavenumber resolution.
In contrast, time-distance spatial resolution may get down to 
the wavelength of the waves being used in the analysis.

Woodard's statistical waveform analysis has been hailed as using more
completely than any of the other local helioseismic methods 
the information in the observed wavefield, and therefore having the 
potential to provide more detail about the underlying subphotospheric
conditions. Yet at present the technique is certainly not well 
developed, there is the issue of whether it would 
require impractical amounts of computational resource and 
how much additional information can be extracted compared to 
other methods. 

\section{Dopplergrams}
The main observational signal used in helioseismology is the Doppler velocity, which is obtained by combining intensity measurements in the core and wings of a photospheric absorption line (e.g. Toussaint, Harvey \& Hubbard 1995; Scherrer et al. 1995). Solar intensity images, or filtergrams, are recorded with a CCD camera over short integration intervals to provide one Doppler velocity image every minute. The study of solar oscillations requires long time series of Dopplergrams -- sampled images of the solar line-of-sight velocity convolved by the point-spread function (PSF) of the telescope.

Several instrumental effects complicate the interpretation of Doppler velocities and lead to systematic errors in the seismic measurements. Understanding these effects is an important and challenging task. \cite{Korzennik2004} have discussed imperfections in the MDI image plate scale, the image distorsion, and temporal changes in the optical system. The MDI PSF, which has been estimated by \cite{Tarbell1996} near disk center, is not axisymmetric, implying a different sensitivity to waves propagating in different directions. Woodard et al. (2001) stress that a good knowledge of the PSF is critical in order to be able
to infer by deconvolution the true energy of solar oscillations versus horizontal wavenumber. 

Systematic variations across the field of view are, of course, not only due to instrumental effects but also to purely geometrical effects including foreshortening -- the reduced sensitivity to short wavelengths away from disk center in the center-to-limb direction -- and the projection of wave motion (mostly vertical) onto the line of sight.  In addition, an error in the orientation of the image will affect the mapping into the heliographic coordinate system. Indeed \cite{Giles1999} showed that a $0.1^\circ$ $P$-angle error can lead to a measurable $\sim 5$~m/s spurious flow in the north-south direction caused by a leakage of the 2~km/s solar rotation.  \cite{Giles1999} also found a one-year periodicity in the MDI travel-time measurements, which he attributed to an error in the determination of the direction of the solar rotation axis. The high-precision alignment and merging of Dopplergrams from the GONG ground-based network of telescopes is discussed by \cite{Toner2001}.

A good calibration of the velocity signal requires a physical model of the absorption line profile from which it is derived. Structural inhomogeneities in the solar atmosphere, especially magnetic field concentrations, affect the shape and formation height of a spectral line. \cite{Wachter2006} and \cite{Rajaguru2006} have proposed corrections to the standard calibration to account for systematic errors due to spectral line shape changes in solar active regions. These studies are important in order to make a distinction between a biased estimate of the Doppler signal and a physical perturbation to the seismic wave field. A source of concern is the possibility of a cross-talk between the measured velocity and magnetic signals.  Besides the magnetic field, turbulent flows also affect line profiles. A radiative transfer calculation of spectral line profiles formed at two different heights in the solar atmosphere is presented in this volume by \cite{Haberreiter2007}. 

The observational systematic effects mentioned above have a direct influence on the interpretation of local helioseismological 
measurements. Correcting for all these effects is a formidable, but highly desirable enterprise.

\section{Travel times}
In local helioseismology, the governing equations of solar oscillations are often written in plane-parallel geometry. This approximation is typically valid over a small fraction of the solar disk for horizontal wavelengths that are much smaller than the solar radius.  Ignoring most of the systematic errors discussed in the previous section, we may write the Doppler signal used in helioseismology as the line-of-sight component of the wave velocity,
\begin{equation}
\phi(\bx,t) = {\rm PSF} \star [ \los \cdot \partial_t \bxi(\bx, z_0,t) ] ,
\label{eq.phi}
\end{equation}
where $\bx$ is a horizontal position vector on the solar surface (defined by height $z_0$), $t$ is time, $\los$ is a unit vector pointing in the direction of the observer, and $\bxi$ is the wave displacement vector. Convolution by the PSF is included. It is customary to consider the data cube $\phi(\bx,t)$ in a reference frame which is co-rotating with the Sun to remove the main component of rotation. 

In time-distance helioseismology, the oscillation signal is filtered in Fourier space to obtain 
\begin{equation}
\psi(\bk,\omega) = F(\bk,\omega) \phi(\bk,\omega) ,
\label{eq.prefilter}
\end{equation}
where $\bk$ is the horizontal wave vector, $\omega$ is the angular frequency, and $F$ is a (real) filter function chosen by the observer to remove granulation noise and to select parts of the wave propagation diagram. A standard procedure is to use phase-speed filters \cite{Duvall1997}. This choice is motivated by the   fact that acoustic waves with the same horizontal phase speed $\omega/k$ are confined to the same acoustic cavity  \cite{Bogdan1997}. The notion of optimal filtering remains to be defined, but is appealing \cite{Couvidat2006}.

The fundamental time-distance computation \cite{Duvall1993} is the cross-covariance function between the filtered Doppler velocities at two surface locations $\bx_1$ and $\bx_2$,
\begin{equation}
\label{eq.c}
C(\bx_1, \bx_2, t) = \frac{1}{T-|t|} \int  \psi(\bx_1, t') \psi(\bx_2, t'+t) \; \id t' ,
\end{equation}
where $T$ is the duration of the observation and $\psi$ is the observed signal.  Pre-multiplication by the temporal window function (equal to unity in the interval $[-T/2,T/2]$ and zero outside) is included in the definition of $\psi$. The positive time lags ($t>0$) give information about waves propagating from $\bx_1$ to $\bx_2$, and the negative time lags ($t<0$) give information about waves propagating in the opposite direction. The cross-covariance function provides a means to add random waves constructively; it represents, in some sense, a solar seismogram.  

The goal is to find a three-dimensional model of the Sun which is consistent with a large set of observed cross-covariance functions. Full waveform modelling has not been attempted yet. In practice, only a few parameters are used to describe a cross-covariance function. The most important one is perhaps the travel time.  

Two methods are used to measure travel times from $C(t)$. The first method \cite{Duvall1997} consists of fitting a Gaussian wavelet to each branch of the cross-covariance. For example, the positive-time part of the cross-covariance is fitted with a function of the form
\begin{equation}
A \exp[-\gamma^2 (t-t_{\rm g})^2] \cos[\omega_0(t-\tau_+)] ,
\end{equation}
where all parameters are free. Some spatial averaging is often  necessary before the fit is performed to guarantee convergence. The time $\tau_+$ is called the phase travel time. Similarly a phase travel time $\tau_-$ is measured by fitting a wavelet to the negative-time part of the cross-covariance.    The travel-time differences $\tau_{\rm diff} = \tau_+ - \tau_-$ are sensitive to internal flows, while the mean travel times $\tau_{\rm mean} = (\tau_+ + \tau_-)/2$ are sensitive to wave-speed perturbations (temperature, density, magnetic field). This distinction, however, may be fuzzy due to finite-wavelength effects. Note that the phase travel time is defined modulo $2\pi/\omega_0$.

An alternative definition of travel time was introduced by Gizon \& Birch (2002, 2004) by analogy with a definition used in geophysics \cite{Zhao1998}. This method is designed to measure travel times from cross-covariances measured with short $T$ and with as little spatial averaging as possible. By definition, the travel times $\tau_+$ and $\tau_-$ are given by
\begin{equation}
\tau_\pm  = \int  W_\pm(\bdelta,t) [C(\bx_1,\bx_2,t) - C_0(\bdelta,t) ] \; \id t ,
\label{eq.deftt}
\end{equation}
where the function $C_0$ is a smooth reference cross-correlation derived from a horizontally homogeneous solar model, $\bdelta=\bx_2-\bx_1$, and the
weight functions $W_\pm$ are proportional to $\partial_t C_0$. One advantage of this definition is a straightforward, linear relationship between travel times and cross-covariance functions.

The second definition  of travel time (eq.~[\ref{eq.deftt}]) is useful as it is more robust to noise than the wavelet fits \cite{Roth2007}. It is also much easier to interpret in terms of perturbations to a reference solar model (sec~\ref{sec.forward}).
On the other hand, a wavelet fit returns additional information: the amplitude, $A$, the group travel time, $t_{\rm g}$, the frequency, $\omega_0$, and a measure of the width of the wavelet, $1/\gamma$. Amplitudes, in particular, are known to vary very significantly in the vicinity of active regions \cite{Jensen2006} since  sunspots `absorb' incoming acoustic energy \cite{Braun 1987}. \cite{Woodard2002} and \cite{Rajaguru2006a} provide two different reasons why amplitude variations and travel-time shifts are not independent parameters as a result of inhomogeneous wave damping and excitation.

The main source of random noise in local helioseismology is, by far, the stochastic nature of solar oscillations. A good understanding of the properties of noise is needed  for a correct interpretation of the measurements and, in particular, for solving the inverse problem (see sect.~\ref{sec.inversion}). Noise is specified by the covariance matrix of the travel times, which may be estimated directly by spatial averaging over many samples of the data \cite{Jensen2003}. We note that the noise covariance matrix depends on the fourth moments of the wavefield,
\begin{equation}
{\rm Cov}[ \phi^*(\bk_1,\omega_1) \phi(\bk_2,\omega_2), \phi^*(\bk_3,\omega_3) \phi(\bk_4,\omega_4)] .
\end{equation}
It may be written down explicitly (Gizon \& Birch 2004) when the $\phi(\bk,\omega)$ are assumed to be normally distributed and uncorrelated in Fourier space, i.e. under the assumption that the random function $\phi(\bx,t)$ is spatially homogeneous. This last assumption, however, is not valid near active regions: noise in travel-time measurements also depends on the local solar properties that we are trying to infer from the travel times (this has been happily ignored).

To summarize this section, the challenge is to extract unbiased information from the wave field, which is relevant  and can be interpreted. This can hardly be achieved  without some understanding of the physics of wave propagation through complex media, which we now discuss.

\section{Forward modelling}
\label{sec.forward}
Forward modeling is the process of computing the wave field given a prescribed solar model. The solar interior is a complex medium in at least two different ways. The notion of complexity depends somewhat on which waves we are talking about. First, the solar convection zone has inhomogeneities that are smaller than the wavelengths of solar oscillations and/or that vary faster than the wave periods. For such a medium, ray acoustics is not valid.  The second kind of complexity is that some inhomogeneities, like sunspots, are not small-amplitude perturbations to the average medium,  so that single-scattering approximations cannot be used.   In the case of turbulent convection, both levels of complexity are combined.

Needless to say that a lot of work remains to be done in order to model wave propagation through complex/random media in the context of local helioseismology. For the sake of simplicity, it is often assumed that there exists a separation of scales between turbulent convection, waves, and slowly-varying structures.  The effects of turbulent flows are twofold: excitation and damping of the waves. Both effects are usually parametrized using phenomenological models. Nearly all studies have assumed that the solar waves ($\sim 5$~min periods) propagate in a background that is steady over the duration, $T$, of the observations (typically several hours).

So far, most efforts have focussed on wave propagation through a steady, weakly inhomogeneous background specified by small-amplitude perturbations to a reference solar model invariant under horizontal translation. This class of problems has been studied extensively in geoseismology. The basic idea is to use first-order perturbation theory -- the first Born or Rytov approximations -- to write the scattered wave field in terms of the zero-order wave field.

The equations of solar oscillations (see e.g. Cameron, Gizon \& Daiffallah 2007) may be combined in the form \cite{LyndenBell1967}
\begin{equation}
L[\bxi] + \Upsilon [\bxi] = \bS , 
\end{equation}
where $L = \rho \partial_t^2 +\cdots$ is a linear differential operator governing adiabatic oscillations,  $\Upsilon$ is a phenomenological damping operator, and $\bS$ is a stochastic source function (granulation).  Using subscripts $0$ to label the unperturbed quantities defined in the reference solar model, we have  $(L_0 + \Upsilon_0) [\bxi_0] = \bS_0$. The background solar model is assumed to have no magnetic field and no flow. The random source function, $\bS_0$, is assumed to be horizontally homogeneous and stationary in time. For a background model where density and pressure only depend on height, $\bxi_0$ can relatively easily be written in terms of a Green's tensor using a normal mode summation (Birch, Kosovichev \& Duvall 2004). In the first-order Born approximation,  the first order perturbation to the wave displacement, $\delta\bxi$, is such that
\begin{equation}
(L_0 + \Upsilon_0)[\delta\bxi] = - (\delta L + \delta\Upsilon)[\bxi_0]  + \delta\bS .
\label{eq.born}
\end{equation}
Inhomogeneities, for example sound speed variations $\delta c(\br)$ or a buried magnetic field $\bB(\br)$, are encapsulated in the wave operator $\delta L$. Note that we have included the perturbations to the damping operator and to the source function. These two terms should not be neglected a priori. For example, it is not excluded that a $\delta \bS$  could arise from a change in the properties of granulation due to a local change in sound-speed (or temperature). These terms, though, are unlikely to be dominant, except perhaps in magnetic regions (see previous section).  The unperturbed Green's tensor can be used to solve equation~(\ref{eq.born}) for $\delta\bxi$, just like it was used to solve the zero-order problem. The Born approximation is an equivalent-source description of wave interaction (the right hand side of eq.~[\ref{eq.born}]). It has been used to compute the effect of sound speed perturbations (Birch, Kosovichev \& Duvall 2004) with
\begin{equation}
- \delta L [\bxi_0]  = \bnabla( \rho_0 \delta c^2 \bnabla\cdot\bxi_0)  .
\end{equation}
In this volume \cite{Birch2007} consider, for the first time, the effect of a flow, $\bU(\br)$, on wave travel times using
\begin{equation}
- \delta L [\bxi_0]  \simeq  2  \rho_0 \bU \cdot \bnabla \partial_t\bxi_0 .
\label{eq.bornc}
\end{equation}
In the case of magnetic perturbations, the appropriate operator to consider  (Lorentz force) is 
\begin{equation}
- \delta L [\bxi_0]  =  \frac{1}{4\pi}( \bJ' \times \bB + \bJ\times \bB')
\label{eq.bornB}
\end{equation}
where currents are given by  $\bJ = \bnabla\times\bB$ and $\bJ' = \bnabla\times\bB'$ and the magnetic field oscillations are $\bB' = \bnabla\times(\bxi_0 \times \bB)$. Note that to first order the perturbation is quadratic in the magnetic field. 

Once the wave displacement, $\bxi = \bxi_0 + \delta\bxi$, has been computed, it is possible to obtain the first-order perturbation to the cross-covariance, $\delta C = C-C_0$, using equations~(\ref{eq.phi})-(\ref{eq.c}). The next step is to relate the observations, in particular the travel times, to $\delta C$. This operation is trivial with the one-parameter fit since $\delta\tau = \int W \, \delta C \, \id t$. It is much less obvious with the (non-linear) wavelet fit. 

The result of single-scattering theories is that there exists a linear relationship between travel-time perturbations and small-amplitude changes in internal solar properties. It is helpful to cast the observational constraints into the form
\begin{equation}
\delta\tau_i = \sum_\alpha\int_\odot 
K_i^\alpha(\ru)
\delta q_\alpha(\ru) 
\, \id\ru   ,
\label{eq.tauq}  
\end{equation}
where $i$ indexes the $M$ different measured travel times (so $i$ runs from $1$ to $M$). The index $\alpha$ refers to the various types of physical quantities, $q_\alpha$,  like sound speed, density, first adiabatic exponent, flows, magnetic field, etc.  The functions $K_i^\alpha(\ru)$ give the sensitivity of small changes in travel time, $\delta\tau_i$, due to a change $\delta q_\alpha(\ru)$ throughout the solar volume.

The accuracy of the Born approximation for magnetic perturbations has been tested by \cite{Gizon2006} using the exact solution for waves impacting a magnetic cylinder in an otherwise uniform medium. For a one kilogauss magnetic field, the Born approximation would appear to be valid except close to the solar surface (the first few 100~km). The assumption of small perturbations breaks down in active region sub-photospheres. At this point a treatment of the strong perturbation regime is needed, a major topic of current research. A very good account of the types of problems encountered in the magnetoseismology of active regions is provided in this volume by \cite{Cally2007} using an extension of ray theory to magnetoacoustic waves. We refer the reader to this paper and references therein.

\section{Inversions}
\label{sec.inversion} The linear inverse problem consists
of making inferences about the unknown functions 
$\delta q_\alpha(\ru)$ from the measurements $\delta\tau_i$.
We assume (see previous section) that the kernel functions
$K_i^\alpha(\ru)$ are known sufficiently accurately.

Equation~(\ref{eq.tauq}) 
can be written in a vector notation, with a dot product 
defined in an obvious way, as
\begin{equation}
\du = \int_\odot 
\Ku \cdot \dqu
\,\id\ru
\end{equation}
or equivalently
\begin{equation}
d_i = \int_\odot 
\Ku_i \cdot \dqu
\,\id\ru  .
\end{equation}

In practice, all the data analysis is done on computers where 
quantities are represented in discretized form. Even if that
were not so, various methods of solving the inverse problem involve
making a discretized representation over
a finite basis of functions. Hence we write 
\begin{equation}
\du = A\xu
\end{equation}
or
\begin{equation}
d_i = A_{ij} x_j  ,
\end{equation}
where $d_i$ represent the data, $x_j$ is a discrete representation of
the unknown functions, and multiplication by $A_{ij}$ replaces the 
integral with weighting given by the kernel functions. Let us say that
$j$ runs from $1$ to $N$.

Taking a linear combination of the data constraints, weighted by some
as yet arbitrary coefficients $c_i$, one obtains
\begin{equation}
\sum_i c_i d_i = 
\int_\odot 
\left(\sum_i c_i \Ku_i\right) \cdot \dqu  
\,\id\ru  .
\end{equation}
The combination $\CKu\equiv\sum c_i\Ku_i$ is called the {\sl averaging
kernel}.  Typically one attempts to localize $\CKu$ near some target
location $\ru_0$; in the case that there
are several components $\alpha$, localization means localizing the 
component corresponding to one chosen $\alpha$ whilst simultaneously making
the components of $\CKu$ corresponding to all other $\alpha$s small.
There are various ways to choose the coefficients $c_i$, different ways
leading to different `solutions' of the inverse problem. The set of such methods forms a family of optimally localized averaging (OLA) methods. An OLA procedure for the inversion of f-mode time-distance helioseismology is given by Jackiewicz et al. (2007). 

A very different approach, at least philosophically, is `data fitting',
by finding the solution $x_i$ to minimize the data mismatch
\begin{equation}
{\rm min\ } \|\du - A\xu\|^2   .
\label{eq.min}
\end{equation}
In practice, strictly minimizing the data mismatch typically leads to 
an unacceptable solution $\xu$ for noisy data, because matrix $A$ is
usually ill-conditioned. The problem must be {\sl regularized} in 
some way. One way that is common in global helioseismology but less so
in local helioseismology is so-called Tikhonov regularization, in which
one minimizes not simply the data mismatch but the sum of this plus a 
penalty term that gets large as $\xu$ exhibits some undesirable property
such as having a large magnitude or large fluctuations. Another way is to 
solve the minimization problem~(\ref{eq.min}) in a succession of iterative
steps that would eventually converge to the exact solution, but instead
truncate the process after a small number of iterations before the 
exact solution has been reached and, more importantly, before the 
solution in attempting to fit closely the noisy data has developed
unacceptable features. Such a method is the so-called
LSQR \cite{Paige1982}.

Complementary to the inversion methods above is so-called
forward-modeling, in which one searches through a space of possible
models to find a best fit to the observational data. One such search
method is genetic modeling, which has been used in global helioseismology
by \cite{Charbonneau1998} and recently in local helioseismology of 
sunspots by \cite{Crouch2005}. 

One consideration in picking an inversion method is how long the
solution will take to compute, or indeed whether the problem will even
fit within the capacity of the available computers. Typically, solving
an OLA problem involves inverting one or more matrices of size 
$M\times M$. The data-fitting techniques involve inverting one or
more matrices of size $M\times N$. If $M$ and $N$ are very large
but one is much larger than the other, the difference between these
two matrix inversion problems may be a significant practical consideration.

Time-distance helioseismology typically involves taking measurements 
of travel times over a variety of distances, at many locations on the 
visible disk of the Sun. If the kernels are invariant under horizontal
translation of the measurement point, then this can be exploited to 
speed up the inversion. A way of doing this is multichannel deconvolution
(MCD) which was introduced into helioseismology by B.~H.~Jacobsen and 
J.~M.~Jensen \cite{Jensen1998}. After Fourier-transforming the problem in the horizontal,
this reduces the problem to a series of one-dimensional inverse problems
in the depth direction. A refinement to control regularization in the 
horizontal direction has been explored by \cite{Zharkov2006}.

Unfortunately, whilst it is approximately the case that the problem 
is invariant under horizontal translation near the centre of the disk,
foreshortening and line-of-sight effects mean that this assumption 
breaks down as one moves away from disk centre. Kernels for time-distance
helioseismology using f-mode data have been 
calculated by \cite{Jackiewicz2006}, 
demonstrating explicitly that these vary with horizontal location on 
the disk. The kernels can be expressed as a weighted sum of a number of 
different contributions: each of the individual contributions is the same 
at all locations but the weight given to each depends on location. Whether
the computational efficiencies of MCD can somehow be retained when the
kernels are no longer translationally invariant is an open question.

A similar concern arises if the initial background model is 
horizontally inhomogeneous. This would be the case, for example, if
one were to proceed with the inversion iteratively, updating the background
model to incorporate the current inversion results after each linear
inversion step.

In common with global helioseismic studies, an issue in local
helioseismology is how to test the reliability of the inferences
drawn from inversions. The inversion solution depends on the method
chosen and the values selected for various trade-off parameters. These
affect the resolution attained and how errors propagate from the
data to the solution. 
For a strictly linear problem and linear 
inversion method, the resolution can be understood completely in 
terms of averaging kernels: the solution at each point can 
be rigorously interpreted as a convolution of the true underlying
solution with the averaging kernel \cite{JCD1990}. 
Resolution measures can be 
defined based on properties of the averaging kernels. Also, for a linear
inversion method, if the statistics of the data errors are completely
known, then the statistics of the errors on the solution can 
be determined precisely, including the correlation between
errors at different points in the solution. 
Complicating factors that are less easy to 
account for are: systematic errors in the data; inadequacies in the 
description of the data errors; nonlinearity of the 
underlying problem; inaccuracies of the forward model (as expressed in
the kernels), which point could include the issue of nonlinearity. 
The effects of some of these might be tested by making inversions of 
artificial data generated from more or less sophisticated numerical
simulations of waves in models of the whole or part of the Sun.

Another issue is that many different physical effects enter 
(as indexed by $\alpha$ in eq.~[\ref{eq.tauq}]). To 
what extent can these be separated by the inversion, either in 
principle, or in practice given a finite number of noisy data?
As discussed by \cite{Zweibel1995}  in the global helioseismic
context, it is not possible for example to distinguish unambiguously 
between thermal perturbations and magnetic fields with a finite
number of frequency measurements. This is another aspect of the 
ill-posedness of the inversion problem.  Other a priori knowledge,
or assumptions or prejudices, might be introduced to break the 
ambiguity. It may also be possible to supplement the seismic data
with other observational constraints, such as photospheric magnetic
field measurements for instance.


\section{Discussion}
We have tried to provide a short description of some of the current topics of research in time-distance helioseismology. Although much work remains to be done, the general procedure to be followed to infer small-amplitude, steady perturbations in the solar atmosphere is, in principle,  relatively well understood.  The treatment of strong, complex perturbations is, however, still in its infancy. As discussed by Werne, Birch \& Julien (2004), the study of wave propagation through magnetic active regions would benefit a lot from numerical simulations. Indeed large-box realistic simulations of fully-compressible non-linear magnetoconvection are becoming feasible (e.g. Georgobiani et al. 2006; Steiner 2007) offering test beds to validate the various methods of local helioseismology (Zhao et al. 2006). Other tests will be performed using codes that propagate linear waves through inhomogeneous background solar models (e.g. Cameron, Gizon \& Daiffallah 2007;  Hanasoge \& Duvall 2007), which are much less computer intensive. 

Much progress in local helioseismology is to be expected
in the next few years thanks to numerical modelling of wave propagation together with insight gained from asymptotic methods (e.g. Cally 2007; Gordovskyy \& Jain 2007; Gough 2007) and from other fields of physics (e.g. seismology, ocean acoustics). Refined methods of data analysis, such as e.g. the three-point correlation technique of \cite{Pijpers2007}, may also prove very useful in extracting all the information from the MDI/SOHO and the upcoming HMI/SDO observations.

\begin{acknowledgements}
The authors acknowledge support from the HELAS network for their participation
in a workshop in Nice, which was the genesis for this paper. HELAS is 
funded by the European Commission under Framework 6.
\end{acknowledgements}

\end{document}